\documentclass[12pt,a4paper]{article}
\usepackage[usenames,dvipsnames,svgnames,table]{xcolor}                 
\usepackage[T1]{fontenc}
\usepackage{lmodern}
\usepackage{textcomp}
\usepackage[english]{babel}
\usepackage{booktabs}
\usepackage{graphicx}
\usepackage[toc,page]{appendix}
\usepackage{indentfirst, comment}
\usepackage{subfigure}
\usepackage[colorlinks=true,linkcolor=blue,citecolor=blue,urlcolor=blue]{hyperref}
\usepackage{authblk}
\usepackage{appendix}
\usepackage[ruled, vlined]{algorithm2e}
\usepackage{multirow}
\usepackage{wrapfig}
\usepackage{float}
\usepackage{booktabs}
\usepackage{natbib}
\usepackage{fullpage}
\usepackage{pgfplots}
\usepackage{transparent}
\usepackage{enumitem}
\usepackage{subfigure}
\usepackage{amsthm,amsfonts,amsmath,amstext,amssymb,amsopn}

\usepackage{tikz}

\usetikzlibrary{mindmap,backgrounds}

\usepackage{setspace}
\usetikzlibrary{arrows}
\usepackage[font=small,labelfont=bf,labelsep=period,tableposition=top]{caption}
\usepackage{natbib}
\usepackage{rotating}
\usepackage{eso-pic}
\usetikzlibrary{spy}
\usepackage{titlesec}
\pgfplotsset{compat=1.12}
\usetikzlibrary{shadows,positioning}
\colorlet{colD}{red!40}
\colorlet{colIP}{cyan!40}
\colorlet{colV}{blue!40}
\colorlet{colBorder}{gray!70}
\tikzset
  {mybox/.style=
    {rectangle,rounded corners,drop shadow,minimum height=1cm,
     minimum width=2cm,align=center,fill=#1,draw=colBorder,line width=1pt
    },
   myarrow/.style=
    {draw=#1,line width=3pt,-stealth,rounded corners
    },
   mylabel/.style={text=#1}
  }
\usetikzlibrary{automata,positioning}
\title{Bridging the user equilibrium and the system optimum in static traffic assignment: how the cooperation among drivers can solve the congestion problem in city networks}
\author{Valentina Morandi}
\affil{Free University of Bozen - Bolzano, Faculty of Science and Technology, Bolzano (Italy)}

\begin{document}

\maketitle

\begin{abstract}
Solving the road congestion problem is one of the most pressing issues in modern cities since it causes time wasting, pollution, higher industrial costs and huge road maintenance costs. 
Advances in ITS technologies and the advent of autonomous vehicles are changing mobility dramatically. They enable the implementation of a coordination mechanism, called coordinated traffic assignment, among the sat-nav devices aiming at assigning paths to drivers to eliminate congestion and to reduce the total travel time in traffic networks.
Among possible congestion avoidance methods, coordinated traffic assignment is a valuable choice since it does not involve huge investments to expand the road network. 
Traffic assignments are traditionally devoted to two main perspectives on which the well-known Wardropian principles are inspired: the user equilibrium and the system optimum.
User equilibrium is a user-driven traffic assignment in which each user chooses the most convenient path selfishly. 
It guarantees that fairness among users is respected since, when the equilibrium is reached, all users sharing the same origin and destination will experience the same travel time.
The main drawback in a user equilibrium is that the system total travel time is not minimized and, hence, the so-called Price of Anarchy is paid.
On the other hand, the system optimum is an efficient system-wide traffic assignment in which drivers are routed on the network in such a way the total travel time is minimized but users might experience travel times that are higher than the other users travelling from the same origin to the same destination affecting the compliance. 
Thus, drawbacks in implementing one of the two assignments can be overcome by hybridizing the two approaches aiming at bridging users' fairness to system-wide efficiency.
In the last decades, a significant number of attempts have been done to bridge fairness among users and system efficiency in traffic assignments.
The survey reviews the state-of-the-art of these trade-off approaches.
\end{abstract}

\maketitle

\section{Introduction}
\label{sec:introduction}
Road congestion is becoming a serious and one of the most urgent problems in metropolitan areas where the traffic demand is steadily growing. Congestion is a significant burden in terms of wasted time, pollution, industrial costs and road maintenance and alleviating traffic volumes will become more and more urgent as the population grows. 
Travellers complain about traffic congestion because it adds to their travel times a delay that can be used for other activities. On the industry side, delays reduce productivity and, consequently, increase the operating costs. Congestion can influence a lot of economic decisions because it affects the choice of the living place, the working place and the travelling mode for most of the population of a certain area. In addition, congestion continues to increase because of the growing population and the increased motorization of the population. 
According to a new report by INRIX and Centre for Economics and Business Research (CEBR), the annual cost of traffic congestion and gridlocks on individual households and national economies in the U.S., U.K., France and Germany will rise to \$293 billion dollars in 2030 with a cumulative value of congestion cost, from now to 2030, near \$4.4 trillion.
In some cases the road network has been extended to accommodate the growing demand but this is not always convenient in terms of costs/benefits trade-off. 
Regulating congestion without extending the current road network can be done using Intelligent Transportation Systems (ITS), such as ramp metering, reversible lanes, limited access roads, bus lanes, carpooling lanes, express toll lanes, congestion pricing mechanisms, variable message signs, etc., or coordinating traffic assignment. 
Latest developments in autonomous vehicles and vehicle to vehicle communications  are paving the way to the coordination among vehicles and traffic assignment will become soon the ruler of the roost among ITS approaches in eliminating traffic congestion and, more  in general, of reducing the total travel time in traffic networks.
Reviews on congestion reduction methods can be found in \cite{papageorgiou2000freeway}, \cite{peeta2000content}, \cite{sichitiu2008inter} and \cite{luo2004survey}.
In most cases the use of ITS technologies and/or traffic limitations has only partially solved the congestion problem and the result is simply a shift of the congestion in other parts of the city network. This is because even the most efficient sat-nav devices consider only the actual information of the traffic networks without considering the impact of the simultaneous choices on the traffic patterns. In fact, as assessed by \cite{KLEIN2018183}, using real-time sat-nav devices, the resulting traffic pattern is likely to be close to usual inefficient equilibrium rather than near to a system optimal traffic pattern.
A way to overcome this limitation is the traffic coordination, i.e. considering the whole system welfare and assigning paths to users in such a way the congestion problem is solved or, at least, kept at a minimum level. 
Traffic coordination may become an even more powerful tool to reduce congestion in view of a massive use of the autonomous vehicles in the near future. In fact, the traffic coordination has been acknowledge in \cite{speranza2018trends} as one of the most prominent trends in transportation and logistic. 
A centralized system may optimize the network performance and paths may be assigned to vehicles according to an optimal assignment. 
However, traffic coordination can be applied also on current road networks if individual needs are taken into account. 
It is well known that a centralized system optimizing network performance that assigns paths to user without any consideration about fairness among users will tremendously affect the users compliance to the system.
Thus, coordinated traffic assignment on real road networks has to be efficient from the system perspective but also fair for users.

Traffic assignments are traditionally divided into two main approaches inspired by the well-known Wardropian principles: the user equilibrium and the system optimum.
User equilibrium is a user-driven traffic assignment in which each user chooses the most convenient path selfishly. 
It guarantees that fairness among users is respected since, when the equilibrium is reached, all users sharing the same origin and destination will experience the same travel time.
The main drawback in a user equilibrium is that the total travel time is not minimized. In fact, the inefficiencies produced by the user equilibrium are well known in literature with the name "price of anarchy", i.e. the price the system is willing to pay to let users choose the route on their own.
On the other hand, the system optimum is a system-wide traffic assignment in which drivers are routed on the network in such a way the total travel time is minimized but users might experience travel times that are higher than the other users travelling from the same origin to the same destination. 
As assessed in \cite{KLEIN2018183}, the system optimum is the most efficient assignment while being "unstable" since it is unfair and users could not comply with the guidance prescriptions.
Since there are drawbacks in using one of the two main approaches, some attempts to bridge the users' fairness with an efficient traffic assignment have been developed in the last years. 

To this aim, in this survey the literature bridging the two different perspectives will be explained and deeply discussed along with many open research questions that can tackled in the immediate future.
Before going through the latest and the most important developments in fair and efficient traffic assignments models, the survey will go first through the survey methodology in Section \ref{met} and, then, through  the concept of road congestion and to the main methods in optimization to solve the congestion problem in Section \ref{cong}. Then, in Section \ref{equilibrium}, the most common models used in traffic assignment optimization are shown. The two former sections introduce, to a neophyte in the field, to the main concepts needed to understand and implement traffic assignment models. Then, in Section \ref{const}, the state of the art of approaches bridging the user equilibrium and the system optimum will be thoroughly discussed. Finally, in Section \ref{conc}, conclusions and ideas for future research will be provided.

\section{Survey methodology}\label{met}

In this review, we focus on studies where traffic assignment models are formulated to address both the issue of efficiency and fairness in congested road networks.

With this aim, contributes to the literature have been searched through the main scholar databases for  operations research, transportation and game theory. Keywords used for the search: price of anarchy, Braess' paradox, fair traffic assignment, efficient traffic assignment, constrained system optimum, bounded rational user equilibrium.

Starting from the results of the former keywords, a preliminary set of relevant publications has been selected. Then, references therein have been analyzed searching for articles that were missing in the first search phase. In order to keep the number of relevant publications at a reasonable level, only journal publications, books and seminal proceedings have been selected. Research items are mostly presented in a chronological order from older ones to new advances in the specific field. At the end, the survey accounts for a final set  of 95 selected publications. 

\section{Congestion definitions}\label{cong}

Traffic congestion is the result of the imbalance between the network capacity and the demand for transportation facilities. According to \cite{falcocchio2015road} congestion in transportation occurs when the occupancy of spaces by vehicles or people reaches unacceptable levels of discomfort or delay. Congestion phenomena are divided into two main categories: the \textbf{recurring} and the \textbf{non-recurring} congestion. According to  \cite{falcocchio2015road} and \cite{stopher2004reducing} the recurring congestion is the delay that travellers regularly experiences during certain periods of time (for example, the rush hour or morning commute). The non-recurring congestion is a delay due to not predictable events that disrupt the traffic flow such as car breakdowns, crashes, works in progress and bad weather conditions. But how the congestion is measured? When does the congestion appear? The traffic congestion can be detected comparing the actual speed with a theoretic free-flow speed or comparing the amount of vehicles on a certain road segment with a threshold defining the maximum amount of vehicles for which the road segment is considered congestion-free. Congestion intensity measures are many and they allow to understand the level of discomfort that can be experienced on a certain road network. For instance, one intensity measure is the congestion delay rate which is the difference between actual travel time rate and free-flow travel time rate (min/km). The USA Transportation Research Board \cite{Ryus201145} uses the experienced speed in order to classify roads with respect to the Level of Service (LoS), i.e. grades from A (free-flow) to F (forced breakdown congestion). Another intensity measure is the travel time index that is the ratio between the free-flow speed and the experienced speed. The main intensity measures used in traffic assignment problems are two: the road congestion and the user's unfairness. The road congestion is obtained as the ratio of the number of vehicles travelling on the road segment (arc in traffic assignment literature) and the capacity of the road. The road capacity has not to be seen as a strong bound on the number of vehicles that can flow on the road segment but rather It has to be seen as a threshold from which the travel time on the network will start to increase significantly. The users' unfairness is instead measured on the whole path assigned to user and it is a variant of the travel time index. The users' unfairness is defined as the relative difference between the experienced travel time and the free-flow travel time. It depends on the free-flow speed and on the experienced speed but also on the the trip length. According to \cite{falcocchio2015road} longer trips are impacted more by congestion with respect to shorter trip so, considering only the experienced speed, the measure could be misleading. But how to model the travel time on each road considering congestion effects? \cite{stopher2004reducing} pointed out that the congestion is a phenomenon that occurs when the demand exceeds the output capacity. Considering this definition, the underlying assumption is that the travel time and the experienced speed depend on the the demand volume on that road. The relationship between travel time and demand is usually expressed by a so-called latency function, where the travel time is a non linear function of the congestion level expressed as percentage of the capacity saturated by the demand. Related concepts and most used latency functions will be described in \ref{planning}.

When dealing with recurrent congestion, collecting information \cite{ben2013impact} is crucial. In \cite{ben2013impact} an experiment with different level of information accuracy is carried out and the negative effect of low information levels is demonstrated. However, even with full information provided, in case of bottlenecks it is necessary to reconsider network design features. This is the case of ramp metering studies \cite{kachroo2011feedback}. Many cities have developed strategies for congestion reduction and the increase of the public transport use as the congestion charging or the road pricing \cite{de2011traffic}. Road pricing and congestion charging can be developed in several ways \cite{de2011traffic}. The first one is called facility-based and it regards tolling roads, bridges and tunnels only on a few facilities. It can be a single point toll or a distance-based toll. Another pricing scheme is called cordons. Tolls on  cordons are an area-based charging method in which vehicles pay a toll to cross a cordon in the inbound or outbound direction or both. Another toll scheme is the zonal scheme in which vehicles pay a fee to enter or exit a zone or to travel inside the zone. Some other schemes can be implemented using tolls proportional to distance. \cite{de2011traffic} offers some advice on which toll scheme can be chosen depending on the case of study and a good review on congestion pricing technologies is provided. Some works make distinction between congestion charging and road pricing. In \cite{stopher2004reducing} it is considered as congestion charging a situation in which tolls are applied on an area that is most likely congested and as road pricing a situation in which tolls are distance-based. The latter is more fair than the former because tolls are spread along the journey in a progressive way and they depend on how much the travel is long. Also congestion detection is a big issue for people who have to apply some regulations. In recent year many devices have been developed in order to derive vehicle speed, safety distance between vehicles and other congestion parameters. Main methods are RFID sensors, CCTV cameras and vehicle to vehicle communications. 

\section{Traffic assignments and the price for anarchy}\label{planning}

According to \cite{patriksson2015traffic}, transportation planning is usually divided into five steps: goal definition, base year inventory, model analysis, travel forecast and network evaluation. The goal definition step is related to find an agreement on goals and objectives. In the base year inventory step, all the data related to the network and demand patterns has to be collected. In the model analysis is to find the relationship between measured quantities (traffic flows and road congestion, for instance). Model analysis is the result of four different phases: trip generation, trip distribution, modal split and traffic assignment. Trip generation consists in finding the number of trips that originate and terminate in different zone of the studied area. Usually this phase is carried out considering socio-economic, geographic and land use features and the different zones are categorized by main purpose as work, leisure or shopping area. In trip distribution phase some formulas to predict the demand of travellers from an origin zone to a destination zone have to be developed. To an origin zone and to a destination zone is usually associated an OD pair with the demand of transportation from its origin to its destination. Usually the demand of travellers is a function of an attractive parameter for each zone. Modal split is a phase in which we determine the mean of transport used by each traveller. The number of travellers that choose a particular mean of transport depends mainly on travel cost in terms of monetary cost or travel time but sometimes also socio-economic factors affect the choice. Traffic assignment is devoted to assign the demand from an origin to a destination to routes in transportation network. This phase is particularly relevant because an estimate of traffic volumes and travel time is returned. Once the model analysis has been done, a travel forecast is produced using data collected in the goal definition step and re-calibrated with the results of the model analysis step. Finally, the network evaluation is a phase in which alternative transportation network and facilities benefits are evaluated and compared. In this literature review we will focus only on the traffic assignment phase of the model analysis step. 

Traffic assignment is a method that assigns the demand of the OD pair to trips on a transportation network. As input of traffic assignment it is required the OD matrix, representing all the OD pairs with demands, and the network representation (usually a capacitated network). The output is an estimate of the traffic flows on each link and, consequently, an estimation of the travel time on each link. The first attempts to tackle the traffic assignment problem were during a just after the World War II. The first proposed assignment was the so-called \emph{all-or-nothing assignment} as proposed in \cite{campbell1950route}. Since the main assumption of the all-or-nothing assignment is that the travel time does not depend on the flow in the links, all the demand of an OD pair is assigned totally to the shortest path for that OD pair. 

After the traffic research community realised that the all-or-nothing assignment was not realistic, they tried to taking into account congestion effects in routing vehicles. The result is the so-called \emph{latency or link performance function}, i.e. a function in which link travel time depends on the number of vehicles using the link. According to \cite{sheffi1985urban}, in a traffic assignment problem the set of constraint of this problem specifies that the demand of all the OD pair has to be satisfied, the flows has to be non negative, the arc (road) utilization is the sum of all the flows traversing that arc. Each arc $a \in A$ is associated with an link latency function $t_a=t_a(x_a)$ where $x_a$ represent the total flow of vehicles (or the entering rate) on the arc $a$. The latency function is usually assumed convex and non decreasing. The objective function depends on the kind of equilibrium the problem is trying to achieve. The path traversing time depends on the number of vehicles that are flowing through the arcs belonging to the path. The latency on a path is usually defined as the sum over all the arc in the considered path of all the arc latency function value under current conditions. In literature a number of authors have proposed different latency functions. A survey on the used latency function in the literature is proposed in \cite{branston1976link}. Below the most used latency functions are listed:

\begin{itemize}
\item $t=t_0 e^{\frac{x}{c}}$
\item $t=t_0 \alpha^{\beta \frac{x}{c}}$ where $\alpha$ and $\beta$ are parameters .
\item $t=t_0 [ 1+ \alpha  (\frac{x}{c})^\beta] $ where $\alpha$ and $\beta$ are parameters (BPR).
\item $t =
\bigg \{
\begin{array}{rl}
\frac{d}{S_0} & x \geq \delta \\
\frac{d}{S(x)} & x \leq \delta \\
\end{array}
$ 
where $d$ is the distance, $S_0$ is the free-flow speed and $S(x)$ the speed experienced with flow greater than $\delta$. $\delta$ should be considered as the congestion threshold.

\end{itemize}

The most used latency function in literature is the one proposed by the Bureau of Public Roads (BPR) with $\alpha=0.15$,$\beta=4$ and $c$ is the arc capacity. 

In the wake of capturing congestion effects in traffic assignment models, in 1952 the two Wardrop principles on flow distribution have been stated (\cite{wardrop1952road} and \cite{wardrop1952proceedings}). The first one is called \emph{user equilibrium} and it is based on the assumption that all users are in equilibrium, i.e. no one is willing to change its own route since there are no faster routes on the network. The second is called \emph{system optimum} and it is based on the assumption that the total travel time is minimized and all drivers comply with the guidance prescriptions. In \cite{beckmann1956studies} the mathematical models for the traffic assignment have been developed in forms of a convex non-linear optimization problem with linear constraints and in \cite{frank1956algorithm} have been presented an iterative algorithm to solve the quadratic optimization problems. The combination of the Frank-Wolfe algorithm and traffic assignment models leads to a method for solving the traffic assignment problem alternating the all-or-nothing assignment with a line search approach. This method is nowadays used for the solution of some traffic assignment models. 

\subsection{User equilibrium and System optimum}\label{equilibrium}

When all drivers individually decide the route they will use in travelling from origin to destination, there are no drivers that can unilaterally choose another route because all used route from an origin to a destination are characterized by the same average travel time. This is because each driver decides to use the least duration path and, at the end, all routes have the same travelling time. This equilibrium situation is called \emph{user equilibrium}. The user equilibrium have some underlying assumptions: the drivers have a complete information about the available paths and the network flows are stable over time. According to \cite{sheffi1985urban}, the user equilibrium model is the following:

\begin{align}
\nonumber \min \quad& \sum \limits_{(ij) \in A} 	\int_{0}^{x_{ij}} t_{ij}(\omega)\, d\omega\\
& x_{ij} = \sum \limits_{c \in C}\sum \limits_{k \in K_c} a_{ij}^{kc} y_c^k
& \forall(i,j)\in A \label{MODEL:pathbased_1}\\
& d_c = \sum \limits_{k \in K_c} y_c^k & \forall c \in C \label{MODEL:pathbased_2}\\
& x_{ij} \ge 0 & \forall (i,j) \in A \label{MODEL:pathbased_12} \\
& y_c^k \ge 0 & \forall c \in C \quad \forall k \in K_c. \label{MODEL:pathbased_3}
\end{align}

The objective function is the sum over all arcs of the integral between 0 and the link flow of the arc latency function. This expression has no economical meaning but it is only a mathematical construction as stated in \cite{sheffi1985urban}. The road network is represented through a graph $G=(V,A)$ where $A$ is the set of arc/road segments and $V$ is the set of intersection points. Variables $x_{ij}$ represent the flow on each arc $(i,j)$ while $y_c^k$ represent the flow of the OD pair $c$ on path $k$. Constraints \ref{MODEL:pathbased_1} bounds the arc flows $x_{ij}$ to the flow of the paths passing through the arc $(i,j)$ using the incidence matrix $a_{ij}^{kc}$. Constraints \ref{MODEL:pathbased_2} guarantee to the demand satisfaction. In \cite{sheffi1985urban} proof of existence and uniqueness of the user equilibrium are provided. It is provided also a proof of the correspondence between the user equilibrium definition and the proposed model. In \cite{lujak2015route} it is also stated that this equilibrium corresponds to a Nash equilibrium in a game with a large amount of players. 

When all drivers act together in such a way the total travel time is minimized, we are facing a \emph{system optimum} solution. System optimum occurs when the sum of the latency experienced by all the users is minimized. According to \cite{sheffi1985urban}, the system optimum model is the following:

\begin{align}
\nonumber \min \quad& \sum \limits_{(ij) \in A}  x_{ij} t_{ij}(x)\\
& x_{ij} = \sum \limits_{c \in C}\sum \limits_{k \in K_c} a_{ij}^{kc} y_c^k
& \forall(i,j)\in A \label{MODEL:pathbased_5}\\
& d_c = \sum \limits_{k \in K_c} y_c^k & \forall c \in C \label{MODEL:pathbased_6}\\
& x_{ij} \ge 0 & \forall (i,j) \in A \label{MODEL:pathbased_7} \\
& y_c^k \ge 0 & \forall c \in C \quad \forall k \in K_c. \label{MODEL:pathbased_8}
\end{align}

The objective function is the sum over all arcs of the arc latency function multiplied by the flow on the arc. Constraint set and variables are the same of the user equilibrium formulation.

\subsection{The price of anarchy: the gap between UE and SO}

\begin{figure}
\begin{center}
\includegraphics{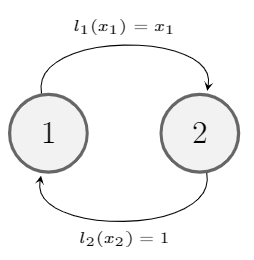}
\end{center}
\caption{Pigou's example}\label{pig}
\end{figure}

Achieving the user equilibrium does not imply that the total travel time is minimized as in the system optimum. In fact, in most cases, as shown in \cite{Harks2015} and in \cite{Jahn2005}, the inefficiency of the equilibrium can be measured. This measure is the \textbf{price of anarchy} and it is defined as the worst-case ratio of the cost of an equilibrium (in terms of total travel time of all the drivers) over the cost under a system-optimum.
An example of the UE with respect to the SO was first provided by Pigou.The Pigou's example network is depicted in Figure \ref{pig} and example description follows. One unit of traffic wants to travel from Node 1 to Node 2 and travel times on the two arcs, upper and lower arc, are depicted in Figure \ref{pig} according to the function $l(x)$ depending on flow $x$. It is easy to see that the travel time from Node 1 to Node 2 on upper arc depends linearly on the flow $x$ while the travel time on the lower arc is always equal to 1, regardless the flow sent on the arc. According to the UE definition, the flow $x$ has to be splitted on the two arcs, with flow $x_1$ and $x_2$, in such a way no portion of the flow is envious, i.e. there no better paths in terms of travel time on the network. The UE equilibrium is attained when the entire flow is route on the upper arc and the total travel time is $x_1 l_1(x_1) + x_2 l_2(x_2)= x_1^2+x_2 1=1^2+0=1$. On the other hand, spitting the flow $x$ in two equal parts, i.e. $x_1=x_2=\frac{1}{2}$, the total travel time is $x_1 l_1(x_1) + x_2 l_2(x_2)= x_1^2+x_2 1=\frac{1}{2}^2+\frac{1}{2}1=\frac{1}{4}+\frac{1}{2}=\frac{3}{4}<1$. In fact, this flow assignment is the so-called SO. Note that this assignment is not valid as UE since travellers on the lower arc are experiencing a travel time which is much higher than the one on the upper arc, in fact doubled. The price of anarchy in the Pigou's example is $\frac{1}{\frac{3}{4}}=\frac{4}{3}$.

Literature on the price of anarchy is wide and bounds has been found for affine and non-negative coefficient polynomial arc latency functions. Considering an instance with latency function $l$ drawn from a family $L$ of non decreasing continuous functions, the price of anarchy is bounded from above by $\alpha(L)$, i.e. $\sum \limits_a x_a^{UE} t_a(x^{UE}) \le \alpha(L) \sum \limits_a x_a^{SO} t_a(x^{SO})$. In a single commodity case and for linear functions the price of anarchy is $\alpha(L)=\frac{4}{3}$, as assessed in \cite{roughgarden2002bad} where bounds for other function families has been also derived. In  fact this is the bound attained in the Pigou's example. In general, for generic function families, the price of anarchy is bounded to be $\alpha(L)=\Theta \left(\frac{p}{\ln p}\right)$  as assessed in \cite{Correa2007}. In \cite{o2016mechanisms} a thorough study on how the demand magnitude impacts on price of anarchy has been conducted. They identified four empirical rules that leads to an increase of the price of anarchy and numerical evidences show that the price of anarchy follows a power law decay for large demands. Most recent findings on the magnitude of the price of anarchy depending on the flow magnitude can be found in in \cite{10.1145/3328526.3329593} and in \cite{colini2020selfish}. The former focuses on bound for the price of anarchy while the latter confirm that the price of anarchy follows a power law with respect to the magnitude of the flow and they stated that, given polynomial latency functions, it can be explicitly evaluated. A micro-simulation framework, embedding features of traffic such as reaction time, acceleration, deceleration, aggressiveness and many others, is provided in \cite{belov2021microsimulation}. The results shows that the price of anarchy embedding such features can be much higher than the theoretical one.

The \textbf{maximum latency price of anarchy} is an alternative way to measure the price of selfish routing. In \cite{Lin2011} the price of selfish routing with respect to the maximum latency experienced by a user is studied. In other words, the user equilibrium total travel time is compared to the total travel time of a min-max latency model. The min-max latency model is a model that minimizes the maximum latency over all experienced paths under the same constraints of the user equilibrium. \cite{Bayram2015146} uses the maximum latency as a measure of unfairness. Bounds on maximum latency price of anarchy bounds have been also derived. In \cite{Correa2007} it is proved that, even for linear latency functions, the maximum latency price of anarchy can be unbounded.  
A further unfairness measure is related to \textbf{Braess's paradox}. In 1968 Braess proposed an example in which the system optimum does not equate with the best overall selfish flow through a network. The Braess's paradox is stated as follows: "For each point of a road network, let there be given the number of cars starting from it, and the destination of the cars. Under these conditions one wishes to estimate the distribution of traffic flow. Whether one street is preferable to another depends not only on the quality of the road, but also on the density of the flow. If every driver takes the path that looks most favourable to him, the resultant running times need not be minimal. Furthermore, it is indicated by an example that an extension of the road network may cause a redistribution of the traffic that results in longer individual running times". 

\begin{figure}
\begin{center}
\includegraphics{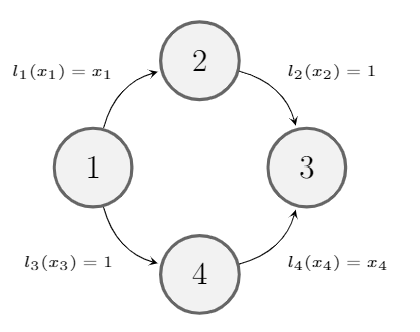} 
\includegraphics{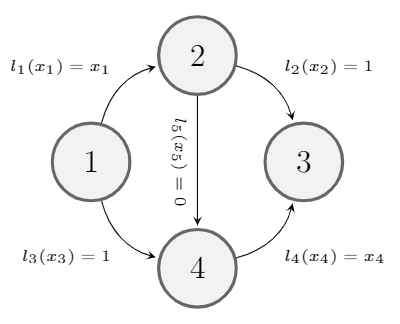} 
\end{center}
\caption{Braess's example - Network before and after}\label{br}
\end{figure}

The Braess networks before and after are, respectively, depicted in Figure \ref{br}. One unit of flow has to be routed from Node 1 to Node 3.The network after differs from the network before only by having added a new arc from Node 2 to Node 4 with constant latency function equal to 0. This change may appear as irrelevant since the travel time on arc is equal to zero. However, the situation dramatically change. In fact, the UE assignment on the before network is halved in the two feasible paths, i.e. 1-2-3 and 1-4-3. This means that the resulting flows are $x_1=x_2=x_3=x_4=\frac{1}{2}$ and the total travel time is $x_1 l_1(x_1)+x_2 l_2(x_2)+x_3 l_3(x_3)+x_4 l_4(x_4)=\frac{1}{2} \frac{1}{2}+\frac{1}{2} 1+ \frac{1}{2} 1+\frac{1}{2}+\frac{1}{2}=\frac{1}{4}+\frac{1}{2}+\frac{1}{2}+\frac{1}{4}=\frac{3}{2}$. Interesting enough, this assignment corresponds to the system optimum and, hence, this network is not affected by the price of anarchy. However, in the after network the former is no longer valid since there exists a new path from Node 1 to Node 2 that results to be cheaper in terms of travel time and so flows are no more in equilibrium. The UE, in this case, is attained when the entire flow is routed on path 1-2-4-3 with an enormous increase in terms of total travel time, $x_1 l_1(x_1)+x_2 l_2(x_2)+x_3 l_3(x_3)+x_4 l_4(x_4)+x_5 l_5(x_5)=1+0+0+1+0=2$. Note that the system optimum on the after network remains the one we had in the before network.

Examples and real world instances of Braess's paradox are shown in \cite{sheffi1985urban} and in \cite{PhysRevLett.101.128701} where such situations have been detected in big cities as London and New York. 

The price of anarchy induced by Braess's paradox is the so-called \textbf{Braess's ratio}. Let $L_i(G)$ is the common latency of the $i-th$ OD pair on a graph $G$. Let $H \subseteq Q$ a subgraph obtained removing arcs from $G$ paying attention in having at least a path for each OD pair. Braess's ratio is: $\beta(G)=\max \limits_{H \subseteq G} \min \limits_{i=1,...,k} \frac{L_i(G)}{L_i(H)}$. In \cite{Lin2011} it is shown that the maximum latency price of anarchy is an upper bound for the Braess's ratio. In \cite{roughgarden2006severity}, a wide study on how to remove Braess' paradox phenomena from networks is provided. they proved that there are no approximation algorithms under a precision threshold to detect and eliminate the paradox and, thus, an efficient detectability is not possible.

Braess' paradox has been studied also with flows over time in \cite{macko2013braess} and according to the traffic assignment with flows over time proposed in \cite{koch2011nash}. They compared with the stationary flow Braess' paradox as the provided example with the one over and have provided examples in which the Braess' paradox appears only when the flow over time is considered. In \cite{akamatsu2003detecting}, the Braess' paradox over time is also studied examinating different example networks and queueing patterns in which the paradox is unavoidable.

Research directions and further insights on the price of anarchy can be found in \cite{10.1007/978-0-387-09680-3_2}.

\section{Bridging the user equilibrium and the system optimum: methodologies and related approaches}

\begin{figure}
\begin{center}
\includegraphics[width=0.4\textwidth]{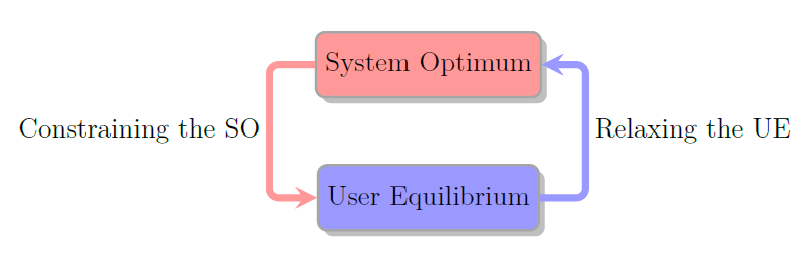}
\end{center}
\caption{Bridging UE/SO scheme}\label{brid}
\end{figure}

\begin{figure}
\begin{center}
\includegraphics{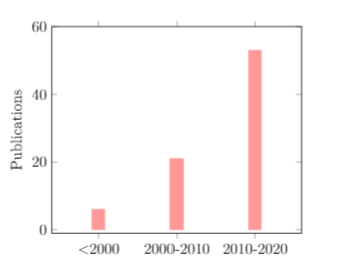}
\end{center}
\caption{Number of contributions through the years}\label{nu}
\end{figure}

As pointed out in the introduction, models bridging the user equilibrium and the system optimum have received great attention during the last years, as assessed by numbers in Figure \ref{nu} where it is clearly shown that the number of publications in the last ten years is considerably increased. The approaches bridging the user equilibrium and the system optimum are mostly divided into two branches: introducing users' fairness constraints in the system optimum and relaxing the requirements of the user equilibrium. As shown in Figure \ref{brid}, from the system optimum we can achieve the user equilibrium by imposing a certain level of fairness among users. On the other hand, the system optimum can be achieved from the user equilibrium by relaxing the fairness requirements imposed by the pure user equilibrium. This review accounts for this distinctions by showing the state-of-the-art of system optimal models with users' constraints in Section \ref{const} and the state-of-the-art of relaxed user equilibrium models in Section \ref{brue}. Furthermore, the Section \ref{stack} is related to the game theory concept of Stackelberg routing and gives the reader pointers to the literature in the field. 

\begin{table}
\label{table}
\setlength{\tabcolsep}{3pt}
\begin{tabular}{|p{25pt}|p{50pt}|p{140pt}|}
\hline
Year& 
Reference& 
Approach \\
\hline
$1999$ & 
\cite{jahn2000optimal} & 
CSO with paths constrained in length \\
$2003$& 
\cite{feldmann2003selfish}& 
Theoretical bounds minimizing the maximum latency \\
$2005$& 
\cite{jahn2005system}& 
CSO with paths constrained using the use equilibrium travel time \\
$2006$& 
\cite{schulz2006efficiency}& 
bounds for the general CSO \\
$2007$& 
\cite{correa2007fast}& 
Variant of the CSO minimizing the maximum latency \\
$2008$& 
\cite{li2008integrated}& 
Integrated equilibrium routing problem\\
$2013$& \cite{Cortes2013313}&
CSO models for large-scale evacuation problems\\
$2015$& 
\cite{lujak2015route}& 
Agent based CSO embedding no envy criteria \\
$2015$& 
\cite{Bayram2015146}& 
CSO in shelter location and evacuation problems\\
$2016$& \cite{bayram2016optimization}&
CSO models for large-scale evacuation problems\\
$2016$& 
\cite{Angelelli2016proactive}& 
Linear hierarchical CSO with paths constrained in length \\
$2018$& 
\cite{Angelelli2018234}& 
Fast large-scale heuristic for \cite{Angelelli2016proactive} \\
$2018$& 
\cite{bayram2018stochastic}& 
CSO in shelter location and evacuation problems under uncertainty \\
$2019$& 
\cite{yuan2019evacuation}& 
CSO in evacuation strategy optimization\\
$2019$& 
\cite{angelelli2019trade}& 
Linear CSO minimizing congestion on the most congested arcs \\
$2019$& 
\cite{van2019travelers}& 
Agent based CSO adding the unfairness marginal cost \\
$2020$& \cite{angelelli2017system} &
Linear CSO with paths constrained using the user equilibrium travel time and large-scale heuristic\\
$2020$& 
\cite{angelelli2019system}& 
CSO bounding real unfairness experienced with loaded arcs \\\hline

\end{tabular}
\caption{From System optimum to User Equilibrium}
\label{tab1}
\end{table}

\subsection{System optimal traffic assignment with users' constraints}\label{const}

The System optimal traffic assignment with users' constraints is a system-optimum traffic assignment problem in which, within the set of constraints, constraints on unfairness among users are introduced int the formulation. These constraints are called side constraints and allow to take into account user-friendly additional restrictions. In fact, the system optimum is difficult to implement for real world networks because it could be very unfair with a subset of system users. As for the pure system optimum, system optimal traffic assignment with users' constraints pays the price of anarchy but it is strongly reduced with respect to the user equilibrium one and it depends the tightness of the side constraints. On the other hand, the system optimum could lead to unacceptably long paths for some drivers while one with side constraints can help in reducing the users' unfairness. It also true that a system optimal traffic assignment with users' constraints, in most cases, is unfair with respect to an user equilibrium which is the fairest assignment. Results on the upper and lower bound on the total travel time, that can be obtained by applying or not cooperation policies, are introduced in \cite{feldmann2003selfish}.

When side constraints are refereed to users' travel time and/or path length, the approach is called constrained system optimum and it is the most prolific research area in the field of system optimal traffic assignment with users' constraints. The idea underlying the constrained system optimum is to propose a little sacrifice in terms of length or travel time to some drivers in order to improve congestion on the whole network. In constrained system optimum, for each OD pairs, a feasible paths set is generated containing only those paths that are not longer/slower than a fixed percentage of the shortest/fastest path for the OD pair. In order to measure how fast a path is, the generating path algorithm uses the so-called normal length. The normal length is an a priori estimate of the real travel time. The path set will be constructed taking only those paths that have normal length shorter than the OD pair shortest path normal length multiplied by a percentage. Usually the arc length, the arc free-flow travel time and the arc travel time under user equilibrium are used as normal length measure. To the best of our knowledge the first attempt to tackle a constrained system optimum is \cite{jahn2000optimal}. In this work traffic flows are routed through a road network in such a way the total road usage is minimized while proposing to users only those paths that are not too long in terms of geographical length as in \cite{mohring2013optimal}. The proposed model is formulated as a non linear multi OD-pair flow problem. They use as latency function the Davidson's function $t_a(x)=t_a^{freeflow}+\frac{\alpha x_a}{u'_a-x_a}$ where $\alpha$ is a tuning parameter and $u'_a$ a parameter chosen in such a way $u'_a>u_a$.  They proposes as solution method the Frank-Wolfe algorithm using, in order to search a feasible direction, a linearization of the non linear problem where eligible paths are generated using a column generation technique. They pointed out that, even in the linearized version, the problem of finding flows on a network in such a way the total travel time is minimized and the followed paths are not longer then a threshold is known to be NP-hard. The impact of choosing a constrained system optimum traffic assignment is widely explained in \cite{Schulz2006} where a theoretical work on the efficiency and fairness is proposed. They measures efficiency by comparing the output of the constrained system optimum with the best solution without guidance and to the user equilibrium while the unfairness is measured comparing travel times of different users. They measure and prove upper bounds of unfairness and efficiency of the constrained system optimum considering different classes of latency functions (affine, non decreasing differentiable, etc.). One interesting features of this work is that they compare these results using as normal length either the free-flow travel travel time and the travel time under user equilibrium conditions. They pointed out that the use of the travel time under user equilibrium conditions as normal length is more reliable a priori estimate of the travel times since they depend also on the traffic volume that intends to travel on the network. According to this modelling choice, \cite{Jahn2005} proposes a constrained system optimum model and methodology that involves using the most commonly used latency function provided from the USA Bureau of Public Road, $t_a(x)=t^{UE}_a[ 1+ 0.15  (\frac{x}{u_a})^4]$ where $t^{UE}_a$ represents the arc travel time under user equilibrium conditions. They propose as measure of the unfairness the comparison between experienced travel times with the best travel times, with the free-flow travel times and with the travel time under user equilibrium. As methodology they propose a variant of the Frank-Wolfe algorithm with a column generation technique in the linearized sub-problem. They provide a wide computational study in which they test the model on seven real road networks where demands are generated using estimations of real data. They shows that using a constrained system optimum the unfairness experienced by users (considering all the unfairness measures) is small. In \cite{Correa2007} a variant of the constrained system optimum with the minimization of the maximum latency is proposed. They propose results on the unfairness and they compare results obtained using either the maximum latency and the total latency as objective value. They show that a flow optimal for the total latency is near-optimal with respect to the minimum maximum latency and it is quite fair and also minimizing with respect to maximum latency produce an optimal solution that is within a constant factor with respect to the optimal solution produced with the total latency. One of the reasons motivating this work is the study of the bottlenecks where a minimum maximum latency level has to be guaranteed in order to avoid the typical phenomena related to bottleneck congestion. 
Theoretical bounds on the objection function value using different latency functions are derived in \cite{Schulz2006} for the traditional constrained system optimum while theoretical bounds when the maximum latency is minimized are provided in \cite{feldmann2003selfish}. 
The first attempt to use a linear programming model to solve the constrained system optimum traffic assignment problem is presented in \cite{Angelelli2016proactive}. In his paper, the proposed approach is hierarchical. First, a linear programming model is run in order to lower the maximum congestion level on the network and, then, a second linear programming model is run to route drivers on fair paths without exceeding the maximum congestion level found in the first model. They shows that it is always possible to lower the maximum congestion level with a lower level of overall experienced unfairness. In details, the second model minimizes the total travel time on selected paths while keeping the network non-congested, if possible, or at its minimum congestion level, otherwise. The set of eligible paths is generated a priori as for \cite{Jahn2005}.
As the number of paths generated a priori is, in the worst case, exponential in the instance size, in \cite{dcvar2017} a heuristic algorithm is proposed. They shows that the algorithm returns a solution which is very near to the optimal one while having very short computational time even on big networks.
In \cite{angelelli2017system} a linear programming model with a traffic-dependent latency function is presented. More specifically, the model makes use of a piecewise linear approximation of the convex latency function. Here, the BPR latency function is embedded in the linear programming formulation. Thorough computational results assess the potentiality of using the linear formulation that allows to take advantage of extremely powerful linear commercial solvers. The linear programming model is able to provide solution in big road networks and, for very big road networks, two heuristic algorithms obtaining excellent results are provided. 
Models analyzed so far are devoted to the minimization of the total travel time and to the minimization of the maximum latency. Both perspectives are of interest but, most of the time, the intermediate shades are the goals a traffic planner aim to achieve. Considering only the maximum latency, the model will not consider the average value of the additional travel time on the road segment and the total travel time could be very bad. On the other end, minimizing the total travel time is totally blind from the point of view of the variability among different road segments. To this aim, in \cite{angelelli2019trade}, a constrained system optimum model able to control the right tail of the distribution of congestion on arcs. They shows that the obtained assignment produces almost the same total travel time of the traditional constrained system optimum model while guaranteeing a very good level of fairness in the spreading of congestion over the arc of the network. 
The constrained system optimum formulations proposed so far aim at minimizing the total travel time while guaranteeing a given fairness level among users but, since eligible paths are generated a priori, the level of the experienced unfairness could turn out to be higher than the a priori given level. Even though in most cases, these eligible paths are the ones involved in the final solution, the choice is not made on the basis of the real flow and, hence, some useful paths can be missed. In order to overcome the drawbacks of the current state-of-the-art traffic assignment models, in \cite{angelelli2019system} two constrained system optimum formulations are provided where the path selection is embedded into the formulations and, thus, the real experienced unfairness is directly controlled inside the model. In \cite{angelelli2019system} the benefits achieved with the new modelling choice are shown and explained also through formal properties.

In \cite{li2008integrated} and in \cite{zhenlong2008integrated} a double objective related to the constrained system optimum achievement is proposed. Literally, it is not a constrained system optimum but a trade-off between system optimum and user equilibrium called integrated-equilibrium routing problem. It calculates the system total travel time under SO, $T_{SO}$, and the travel time for each user under user equilibrium $t_{UE}$. Then, the first objective $Z_1$ is the classic system optimum one and the second objective $Z_2$ is the classic user equilibrium one. Constraints set is as usual plus a constraint on the system optimum $Z_1 \le T_{SO}+ \epsilon$ and a set of constraints, one for each user, $Z_2 \le t_{UE} + \xi$ where $\epsilon$ and $\xi$ are functions respectively of $T_{SO}$ and $t_{UE}$. 

Attempts to bridge the system optimum with the user equilibrium have been done also in field of agent-based models and/or simulations. In \cite{lujak2015route} an agent-based model is proposed that uses a new set of constraints in which the unfairness is bounded by a no-envy criteria between users. \cite{lujak2015route} introduces the concept of normalized mean path duration as the geometric mean of the flow on a path multiplied by the number of driver using that path for each commodity. In mathematical notation, it is $\gamma_c=\sqrt[|P_c|]{\prod \limits_{p \in P_c} f^p x^p}$ where $c$ is the commodity and $P_c$ its path set. The no-envy criteria for each commodity is: $\gamma_c \ge \gamma_{c'}^{\alpha}$ with $0 \le \alpha \le 1$ and $\forall c \in C$, $c' \in C,c'\neq c$. Other examples can be found in \cite{LEVY2017}, in \cite{KLEIN2018183} and in \cite{levy2018emergence} where other agent-based models on route-choice games are presented.
Among agent-based simulations, a social routing framework has been proposed in \cite{van2019travelers} that aims at attaining a pseudo-system optimum traffic assignment through assigning path to users so as to minimize the total travel time plus the marginal cost for compliant users that are routed on paths that are sub-optimal with respect to the fastest path on the network. They divided the possible scenarios into selfish, social and mixed scenarios based on the fraction of the demand which is willing to follow the pro-social instructions and, hence, to follow path that are different from the fastest one. They found out that savings in terms of total travel time are remarkable. In fact, as assessed by \cite{djavadian2014empirical}, some drivers are more willing to follow the guidance instruction in order to enhance the system benefit. Hence, they divided the set of drivers accordingly. 

Besides traffic assignments models, there exist in literature other lines of research that can be easily associated with system optimum with users' constraints. In the following some examples are provided. The constrained system optimum, as proposed in \cite{Jahn2005}, has been successful used also for managing traffic under emergencies and/or natural catastrophes. A constrained system optimum approach has been presented also in \cite{Bayram2015146}, where the model is applied to shelters location and evacuation planning in disasters' management. The approach first assigns users to shelters and, then, users are assigned to the shortest path to their shelter with a given degree of tolerance. The set of feasible paths is determined as in \cite{Jahn2005}. A second order cone programming technique is used to efficiently solve the problem. In \cite{Bayram2015146}  the total evacuation time is minimized under optimal location of the shelters. The constraints are the shelter assignment ones. We recall that the traversing time depends on how many vehicle are standing on an arc so it is not convenient to route all evacuees on the shortest path. A stochastic version of the problem with different scenarios is tackled in \cite{bayram2018stochastic} while a Bender's decomposition approach is proposed in \cite{bayram2018shelter}. In \cite{bayram2016optimization} a very comprehensive review on traffic assignment models for evacuation planning is also provided. In \cite{yuan2019evacuation} a constrained system optimum is proposed to evacuate areas in a secure and stable way. They define the  evacuation time as the time needed from the start of the emergence to reach the evacuee secure area and they constraints the evacuation time of all users to be within a tolerance factor of the fastest user evacuation time.

Paths selection is not always related to the path length or duration. In \cite{Cortes2013313} a set of diversification constraints is provided. These constraints allow to demand to be sent on at least a predefined number of paths or a number of arc-disjointed paths, i.e. paths without arc in common. One particular difference between this work and the others is that the cost function is concave. That is because the problem regards supply chain management and the curve represents scale economies effect in transportation problems. See references for details. An optimal iterative algorithm based on the Kuhn-Tucker optimality conditions is provided. 

On the network design side, the issue of fairness, here called equity, is considered when new infrastructures have to be constructed and the consequent effect on traffic flows have to be evaluated; see \cite{yang1998models}, \cite{patriksson2008applicability}, \cite{LIU201520}, \cite{meng2001equivalent}, \cite{yang2002multiclass} and \cite{GUO2009379} for details and references therein.

Another field of application are the communication networks where the constrained system optimum model is also widely used. We report an example. In \cite{holmberg2003multicommodity} the main issue is to avoid paths with high dispatching delays. The time delay is calculated by summing up the estimated link delay of each arc that belongs to the considered path. For this reason a limit on the cost per unit of flow on each path is calculated. These limits can also include distortion on the network and link failure. A system optimum objective function is used and the constraints set considers only paths which weight is less or equal to the bound due to link delays on that path. The results is a classical constrained system-optimum. Another little difference is in the forcing constraints, i.e. the ones that say if a path is used or not. Since binary variables are used to recognize if a path is used or not, a relaxation is proposed. The problem is solved with a column generation technique.

\subsection{Bounded rational user equilibrium}\label{brue}

The relaxation of the user equilibrium conditions has been studied through years from several perspectives. The relaxation of the user equilibrium conditions goes mainly through limiting the set of feasible path from an origin to destination to those that no longer than the fastest path on the network (constrained user equilibrium), through the definition of a range anxiety due to the needs of vehicles that are routed through the network and through the definition of an indifference band that refer to users' behaviour in perceiving differences among assigned paths, i.e. the so-called bounded rational user equilibrium.

The constrained user equilibrium (briefly CUE) is a user equilibrium in which the path set is limited to the paths that meet some restrictions on length, travel time or other parameters. An example of constrained user equilibrium is provided in \cite{zhou2012user} where the path set is restricted only to those paths that are shorter than the a fixed threshold. They use, in order to compare paths, the path euclidean length and they label as feasible a path only if it is shorter than the shortest path from the origin to the destination multiplied by a scaling factor greater than 1. A path-based formulation of the user equilibrium, in which only feasible paths are allowed, is provided and a column generation technique combined with the Frank-Wolfe algorithm is proposed. They use as latency function the usual one provided by the USA Bureau of Public Road. They propose also a length constrained system optimum formulation with the same path set used for the constrained user equilibrium and they use it in order to formulate an optimal road pricing scheme able to achieve the constrained system optimum solution. These concepts fully apply when electric vehicles are in involved in the traffic assignment. This is because the vehicle battery have its own duration and the trip cannot last more than a fixed threshold. In \cite{jiang2012path} a distance constrained user equilibrium with penalties in the case in which the threshold is exceeded is provided. Further extension related to needs of electric vehicles are provided in \cite{jiang2014computing} and in \cite{jiang2014network}. 

Remaining in the electric vehicles field, the range anxiety phenomena affects the traffic assignment. The range anxiety is the fear of running out of battery en-route and it affects mostly but not only electric vehicles. Given the existence of this phenomena, the resulting traffic assignment could be heavily modified. So, in \cite{he2014network},  \cite{he2015deploying}, \cite{wang2016path} and \cite{xie2017path}, a user equilibrium in which the range anxiety is bounded is presented studying different shades of the problem. The same concept with other kinds of fuel is defined as the relay requirement. The assignment which is the result of bounding the user equilibrium in order to consider the relay requirements is presented in \cite{xie2016relay}. In \cite{xie2019traffic}, a wide up-to-date survey on approaches bounding the user equilibrium to the range anxiety and mental account is provided.

The bounded rational user equilibrium is the main approach that modifies the user equilibrium in order to reduce the price of anarchy and bridge it to the system optimum. The bounded rational user equilibrium was first proposed in \cite{mahmassani1987boundedly} as a relaxation of the user equilibrium in which a path can be used only if its path traversal time is within a range with respect to the fastest path on the network (see \cite{zhang2011behavioral} for further details).
This range is called indifference band and is usually obtained by means of road user behavioral studies or empirical observations. Moreover, this indifference band could be calibrated depending on the OD pair to which is assigned. The concept of indifference band has been successfully embedded in traffic simulations (see \cite{jayakrishnan1994evaluation}, \cite{mahmassani1991system}, \cite{hu1997day} and \cite{mahmassani1999dynamics} for details) as a key elements to bridge the user equilibrium to a more efficient assignment and empirical evidences of the gain in terms of efficiency were described in \cite{jou2005route},\cite{jou2010urban} and later in \cite{di2017indifference}. Mathematical properties and the behaviour of the price of anarchy under bounded rational user equilibrium can be found in \cite{lou2010robust} enlarging the field to optimal tolling strategies. Further mathematical properties are derived in \cite{di2013boundedly}, \cite{di2014braess} and \cite{di2016second}. A complete review on the applications of the bounded rational user equilibrium can be found in \cite{di2016boundedly}. Recent attempts to include the bounded rational user equilibrium into traffic behavioural studies can be found in \cite{ye2017rational}. Although the bounded rational user equilibrium ensures a certain level of fairness among users (considering traffic flows), it suffers from two shortcomings: the equilibrium is not unique (see \cite{zhang2011behavioral}) and it only aims at reaching an equilibrium state without minimizing the total travel time. Thus, no guarantee on the reduction of the price of anarchy can be derived. In fact, the user equilibrium solution is itself feasible for a bounded rational user equilibrium.
A comprehensive review of bounded rational user equilibrium models in a dynamic setting can be found in \cite{szeto2015bounded}.

\begin{table}
\label{table}
\setlength{\tabcolsep}{3pt}
\begin{tabular}{|p{25pt}|p{50pt}|p{140pt}|}
\hline
Year& 
Reference& 
Approach \\
\hline
$1987$& \cite{mahmassani1987boundedly}&
First bounded rational user equilibrium model\\
$2010$& 
\cite{lou2010robust}& 
Congestion pricing under bounded rational user equilibrium \\
$2011$& 
\cite{zhang2011behavioral}& 
User equilibrium under different behaviour assumptions \\
$2012$ & 
\cite{zhou2012user} & 
CUE with paths constrained in length \\
$2012$& 
\cite{jiang2012path}& 
CUE with penalties for paths exceeding the length threshold \\
$2014$& 
\cite{jiang2014computing}& 
CUE with mixed electric and gasoline vehicles \\
$2014$& 
\cite{jiang2014network}& 
CUE with mixed electric and gasoline vehicles and parking slots\\
$2014$& 
\cite{he2014network}& 
UE with range anxiety for electric vehicles \\
$2016$& 
\cite{wang2016path}& 
CUE with range anxiety for electric vehicles\\
$2016$& 
\cite{xie2016relay}& 
CUE with relay requirements for electric vehicles\\
$2016$& 
\cite{di2016boundedly}& 
Review on the bounded rational user equilibrium \\
$2017$& 
\cite{ye2017rational}& 
Adjustment processes for the bounded rational user equilibrium\\
$2017$& \cite{xie2017path}&
CUE with stochastic range anxiety for electric vehicles\\
$2019$& 
\cite{xie2019traffic}& 
Survey on models on range anxiety\\\hline

\end{tabular}
\caption{From User Equilibrium to System Optimum}
\label{tab1}
\end{table}

\subsection{The Stackelberg routing}\label{stack}

In order to decrease the price of anarchy, individuals need some external steering in being cooperative, as they cannot identify socially desired alternatives themselves. To that end, travel information can be quite helpful. For instance, Stackelberg routing (\cite{korilis1997achieving}) assigns a fraction of travellers by a central authority (i.e. leader) as they comply with advice that they received, while the remaining individuals (i.e. followers) choose their route selfishly (\cite{krichene2014stackelberg}). In this,
the leader anticipates on the (expected) selfish response in order to improve overall network performance (\cite{krichene2014stackelberg}). The Stackelberg routing has been proved to be effective, as shown in \cite{bonifaci2010stackelberg} where bounds on price of anarchy have been derived. Several modifications of the pure Stackelberg algorithm have been introduced in the last years as the introduction of tolls (\cite{swamy2012effectiveness}) or the the imperfect knowledge of the duration on certain arcs (\cite{bhaskar2019achieving}).

\begin{figure}
\includegraphics{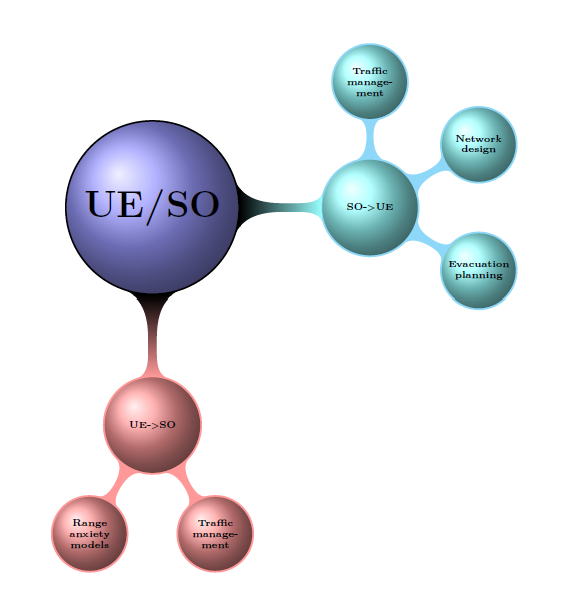}
\caption{Fields of application}\label{fi}
\end{figure}

\section{Conclusions and future research directions}\label{conc}

The survey shows the potentiality of bridging the most well-known Wardrop's principles to efficiently model traffic assignment problems.
This exciting research area has become even more interesting during the last years in which the literature has grown a lot. The proposed literature is surely valuable to traffic planners and it will become more and more interesting with the advent of autonomous vehicles. Many research questions remain open in any branch of the literature and fields of application can be widened to multidisciplinary approaches such as behavioural aspects and developments in ITS technologies. There is also a urgent need for very quick algorithms in order to let these concepts to be embedded into real-time systems especially in view of implementing the 5G technology. As a concluding remark, the survey opens the ideas implemented into traffic assignments to a wider audience. In the era of sharing economy, finding a way to satisfy the users while optimizing the system is no longer only a research question but a need. To this aim, the scope of the survey is also to provide tools to be applied in many other fields. The fields of application are in fact many, as in Figure \ref{fi} where the fields in which this hybrid models have been already applied, but several other fields are facing the dichotomy among efficiency and fairness where the proposed approaches can be applied.

\newpage
\bibliography{Glossary}

\bibliographystyle{natbib}

\end{document}